%Paper: hep-th/9302137
%From: DCROSSLEY@wishep.physics.wisc.edu
%Date: Fri, 26 Feb 93 13:03 CDT

\input phyzzx
%
%%%%%%%%%%%%%%%%%%%%%%%%%%%%%%%%%%%%%%%%%%%%%%%%%%%%%%%%%%%%%%%%%%%%%%%%
%%%                 This is PHYZZX.LOCAL (cm version)                %%%
%%%%%%%%%%%%%%%%%%%%%%%%%%%%%%%%%%%%%%%%%%%%%%%%%%%%%%%%%%%%%%%%%%%%%%%%
%%%%%%%     Latest update/debug: June 24, 1992.         %%%%%%%%%%%%%%%%
%%%%%%%%%%%%%%%%%%%%%%%%%%%%%%%%%%%%%%%%%%%%%%%%%%%%%%%%%%%%%%%%%%%%%%%%
%
\catcode`@=11 % This allows us to modify PLAIN macros.
%
%
%
%	Modifications to PHYZZX to make it UW specific
% 	-includes: a new memo header
% 		   a new letter head
% 		   a UW HEP publication number
%                  a new label output routine for use with LABELFORM.TEX
%
%    ADD SOME MORE FONTS
%\font\twentyrm=cmr10  at 20truept %scaled\magstep4
\newfam\ssfam   % Define a San Serif font family
\font\seventeenss =cmss10 scaled\magstep4
\font\fourteenss  =cmss10 scaled\magstep2
\font\twelvess    =cmss10 scaled\magstep1
\font\tenss       =cmss10
\font\niness      =cmss9
\font\eightss     =cmss8
\def\seventeenpoint{\relax
    \textfont0=\seventeenrm          \scriptfont0=\twelverm
      \scriptscriptfont0=\ninerm
    \textfont1=\seventeeni           \scriptfont1=\twelvei
      \scriptscriptfont1=\ninei
    \textfont2=\seventeensy          \scriptfont2=\twelvesy
      \scriptscriptfont2=\ninesy
    \textfont3=\seventeenex          \scriptfont3=\twelveex
      \scriptscriptfont3=\ninex
    \textfont\itfam=\seventeenit    %\scriptfont\itfam=\twelveit
    \textfont\slfam=\seventeensl    %\scriptfont\slfam=\twelvesl
      \scriptscriptfont\slfam=\ninesl
    \textfont\bffam=\seventeenbf     \scriptfont\bffam=\twelvebf
      \scriptscriptfont\bffam=\ninebf
    \textfont\ttfam=\seventeentt
    \textfont\cpfam=\seventeencp
    \textfont\ssfam=\seventeenss     \scriptfont\ssfam=\twelvess
      \scriptscriptfont\ssfam=\niness
    \samef@nt
    \b@gheight=17pt
    \setbox\strutbox=\hbox{\vrule height 0.85\b@gheight
                                depth 0.35\b@gheight width\z@ }}
\def\fourteenf@nts{\relax
    \textfont0=\fourteenrm          \scriptfont0=\tenrm
      \scriptscriptfont0=\sevenrm
    \textfont1=\fourteeni           \scriptfont1=\teni
      \scriptscriptfont1=\seveni
    \textfont2=\fourteensy          \scriptfont2=\tensy
      \scriptscriptfont2=\sevensy
    \textfont3=\fourteenex          \scriptfont3=\twelveex
      \scriptscriptfont3=\tenex
    \textfont\itfam=\fourteenit     \scriptfont\itfam=\tenit
    \textfont\slfam=\fourteensl     \scriptfont\slfam=\tensl
    \textfont\bffam=\fourteenbf     \scriptfont\bffam=\tenbf
      \scriptscriptfont\bffam=\sevenbf
    \textfont\ttfam=\fourteentt
    \textfont\cpfam=\fourteencp
    \textfont\ssfam=\fourteenss     \scriptfont\ssfam=\tenss
        \scriptscriptfont\ssfam=\sevenrm }
\def\twelvef@nts{\relax
    \textfont0=\twelverm          \scriptfont0=\ninerm
      \scriptscriptfont0=\sixrm
    \textfont1=\twelvei           \scriptfont1=\ninei
      \scriptscriptfont1=\sixi
    \textfont2=\twelvesy           \scriptfont2=\ninesy
      \scriptscriptfont2=\sixsy
    \textfont3=\twelveex          \scriptfont3=\tenex
      \scriptscriptfont3=\tenex
    \textfont\itfam=\twelveit     \scriptfont\itfam=\nineit
    \textfont\slfam=\twelvesl     \scriptfont\slfam=\ninesl
    \textfont\bffam=\twelvebf     \scriptfont\bffam=\ninebf
      \scriptscriptfont\bffam=\sixbf
    \textfont\ttfam=\twelvett
    \textfont\cpfam=\twelvecp    \scriptfont\cpfam=\tencp
    \textfont\ssfam=\twelvess    \scriptfont\ssfam=\niness
      \scriptscriptfont\ssfam=\sixrm }
\def\tenf@nts{\relax
    \textfont0=\tenrm          \scriptfont0=\sevenrm
      \scriptscriptfont0=\fiverm
    \textfont1=\teni           \scriptfont1=\seveni
      \scriptscriptfont1=\fivei
    \textfont2=\tensy          \scriptfont2=\sevensy
      \scriptscriptfont2=\fivesy
    \textfont3=\tenex          \scriptfont3=\tenex
      \scriptscriptfont3=\tenex
    \textfont\itfam=\tenit     \scriptfont\itfam=\seveni  % no \sevenit
    \textfont\slfam=\tensl     \scriptfont\slfam=\sevenrm % no \sevensl
    \textfont\bffam=\tenbf     \scriptfont\bffam=\sevenbf
      \scriptscriptfont\bffam=\fivebf
    \textfont\ttfam=\tentt
    \textfont\cpfam=\tencp
    \textfont\ssfam=\tenss      \scriptfont\ssfam=\eightss
      \scriptscriptfont\ssfam=\fiverm }
\def\ss{\n@expand\f@m\ssfam}
\font\bfss=cmssbx10
\font\fortssbx=cmssbx10 scaled \magstep2
%\font\llogo=uwlogo scaled \magstep0
%
\newdimen\madphheadsize \madphheadsize=7.0in
\newdimen\madphheadleft \madphheadleft=3.75in
\newdimen\madpheadsize \madpheadsize=7.0in
\newdimen\madpheadleft \madpheadleft=3.5in
\showboxbreadth=1000 %
\showboxdepth=5
\newdimen\madphheadsize \madphheadsize=7.0in
\newdimen\madphheadleft \madphheadleft=3.6875in
\def\MADPHHEAD{{\Tenpoint\vbox{\vskip-0.7in
  \line{\hss
   \hbox{\vbox{\hbox to \madphheadsize{\hskip\madphheadleft
                        \fortssbx  University of Wisconsin - Madison\hfil}
               \vskip 5pt
               \hbox to \madphheadsize{\hskip\madphheadleft
                         \bfss Department of Physics\hfil}
               \vskip -45pt
	       \hbox to \madphheadsize{\hskip 3.25in
			{\logo A}
                          \hfill}
               \vskip 24pt \hrule height 1pt \vskip 6pt
               \hbox to \madphheadsize{\hbox to 0pt{\hskip 1pt
                  \tenss High Energy Physics\hss}
    \hskip \madphheadleft
    \tenss 1150 University Avenue, Madison, Wisconsin 53706\hfil}
   \hbox to \madphheadsize{\hbox to 0pt{\hskip 1pt
    \tenss Telephone: 608/262-2281\hss}
    \hskip \madphheadleft
    \tenss Telex: 265452 UOFWISC MDS\hfil}}}\hss}}}}

%%%%%%%%%%%%%%%%%%%%%%%%%%%%%%%%%%%%%%%%%%%%%%%%%%%%%%%%%%%%%%%%%%%%%%%%%%%%%%
%%%%%%%%%%%%%%%%% modify phyzzx for UW label format %%%%%%%%%%%%%%%%%%%%%%%%%%
%%%%%%%%%%%%%%%%  To produce labels, tex LABELFORM.TEX %%%%%%%%%%%%%%%%%%%%%%%
%%%%%%%%%%%%%%%%%%%%%%%%%%%%%%%%%%%%%%%%%%%%%%%%%%%%%%%%%%%%%%%%%%%%%%%%%%%%%%
\begingroup
 \catcode `\{ = 12  % Dirty trick to write out the character {
 \catcode `\} = 12  % Dirty trick to write out the character }
 \catcode `\[ = 1
 \catcode `\] = 2
 \gdef\labelformlabels[%
   \gdef\rwl@begin##1\cr[\rw@toks=[##1]\rel@x
        \immediate\write\labelswrite[\the\rw@toks]\futurelet\n@xt\rwl@next]
   \gdef\writenextlabel##1[%
        \immediate\write\labelswrite[  ]%
        \immediate\write\labelswrite[{]%
            \rwl@begin ##1%
            \rwl@end%
        \immediate\write\labelswrite[}]]%
   \gdef\writelabel##1[%
        \immediate\write\labelswrite[{]%
            \rwl@begin ##1%
            \rwl@end%
        \immediate\write\labelswrite[}]%
        \let\writelabel=\writenextlabel]%
]
\endgroup
%
%\def\addressee#1{\bigskip\medskip
% \line{\hskip 0.5\hsize minus 0.5\hsize \the\date\hfil} \bigskip
% \vskip\lettertopfil
% \ialign to\hsize{\strut ##\hfil\tabskip 0pt plus \hsize \cr #1\crcr}
% \medskip\noindent\hskip\spskip}
%

%\Pubnum={$\caps WISC - EX - 85 - \the\pubnum $}
%%%%%%%%%%%%%%%%%%%%%%%%%%%%%%%%%%%%%%%%%%%%%%%%%%%%%%%%%%%%%%%%%%%%%%%%
%
%\expandafter\ifx\csname eightrm\endcsname\relax
%    \let\eightrm=\ninerm \let\eightbf=\ninebf \fi
%
% some phyzzx overrides
%
\def\figitem#1{\r@fitem{#1.}}
\def\tabitem#1{\r@fitem{#1.}}

\def\sequentialequations{\rel@x \ifnum\equanumber<0 \else
  \gl@bal\equanumber=-\equanumber \gl@bal\advance\equanumber by -1 \fi }

%% a Journal macro that handles NPB and PR format
%\def\Journal#1&#2&#3(#4){\begingroup \let\Journal=\dummyj@urnal
%    \unskip,~#1\unskip~%
%    \ifPhysRev\bf\fi\ignorespaces #2\rm
%    \ifPhysRev\unskip,~\ignorespaces #3\fi
%    \unskip~(\afterassignment\j@ur \count255=#4)
%    \ifPhysRev\else\unskip,~\ignorespaces #3\fi
%    \endgroup\unskip\ignorespaces }
%%
%\def\memohead{\hrule height\z@ \kern -0.5in
%    \line{\quad\fourteenrm LNS MEMORANDUM\hfil \twelverm\the\date\quad}}
%\def\memorule{\par \medskip \hrule height 0.5pt \kern 1.5pt
%   \hrule height 0.5pt \medskip}
%%
%%
%\def\WILSONHEAD{
%  \setbox0=\vbox{
%    \offinterlineskip
%    \hbox{\twentyrm Cornell University\hfil}
%    \vskip 5truept
%    \hbox{\twelvess Floyd R. Newman Laboratory of Nuclear Studies\hfil}
%    \let\myboxwidth=\hsize
%%   \advance\myboxwidth by -35truept % enable if seal done
%    \vskip 3truept\hbox to\myboxwidth
%       {\leaders\hrule height 1truept\hfill}\vskip 4truept
%    \lineskip 3truept
%    \line{\tenss \lnsperson                  \hfil        \lnsuser1}
%    \line{\tenss Wilson Laboratory           \hfil        \lnsuser2}
%    \line{\tenss Cornell University          \hfil        \lnsphone}
%    \line{\tenss Ithaca, NY\quad 14853--8001 \hfil        Telex WUI 6713054}}
%  \setbox2=\vbox to \ht0{\vfil\hbox to 35truept{\hfil~}\vfil}
%% should be to 70pt for seal.  patch to 35pt so looks ok without
%  \medskip\line{\hskip -35truept\box2\box0\hfil}\bigskip}
%\let\letterhead=\WILSONHEAD
%%
%\FromAddress={}
%%
%\def\lnsphone{(607) 255--\lnsext}
%\def\lnsext{4882}
%\def\lnsuser1{}
%\def\lnsuser2{}
%\def\lnsperson{}
%%
\def\boxit#1{\vbox{\hrule\hbox{\vrule\kern3pt
\vbox{\kern3pt#1\kern3pt}\kern3pt\vrule}\hrule}}
%
%
%%%%%%%%%%%%%%%%%%%%%%%%%%%%%%%%%%%%%%%%%%%%%%%%%%%%%%%%%%%%%%%%%%%%%%%%%%%
%%
%%            Making double-column  (these are modified from manmac.tex)
%%            with a full size columns as well.
%%            (This is still buggy--gives overfull boxes etc.)
%%            Report bugs to T J Allen (tjallen@wishep.physics.wisc.edu,
%%            tjallen@suhep.phy.syr.edu or  tja@theory3.caltech.edu)
%%
%%            This will NOT produce double columns in preprintmode since
%%            there are conflicting \output commands.  The whole
%%            macro should be rewritten using a modified \output.
%%
%%            Where you want
%%            the doublecolumn output to start, use \begindoublecolumns.
%%            Where you want to go back to single columns use
%%            \enddoublecolumns.  This produces output much like that
%%            of RevTeX.  If you wish to specify that there be a rule
%%            between the columns of output, then set \columnrulewidth
%%            = 0.4pt.
%%
%%%%%%%%%%%%%%%%%%%%%%%%%%%%%%%%%%%%%%%%%%%%%%%%%%%%%%%%%%%%%%%%%%%%%%
%%
\newbox\partialpage
\newdimen\pageheight \pageheight=\vsize
\newdimen\pagewidth  \pagewidth=6.6truein
\newdimen\columnwidth  \columnwidth=3.2truein
\newdimen\columnrulewidth \columnrulewidth=0pt
\newdimen\ruleht \ruleht=.5pt
\newinsert\margin
\def\twocolumn{%
   \singlespace
   \vsize=9truein
   \pagetextwidth=\pagewidth
   \hsize=\pagewidth
   \titlepagewidth=\pagewidth
   \hoffset=0truein
   \voffset=0truein
   \dimen\margin=\maxdimen
   \count\margin=0 \skip\margin=0pt
 \def\begindoublecolumns{
     \ifpr@printstyle
     \message{ I'm unable to print double columns in PREPRINTSTYLE }
     \end\fi
     \begingroup
     \global\vsize=2\pageheight
     \output={\global\setbox\partialpage=\vbox{\unvbox255\bigskip\bigskip}
         \global\vsize=2\pageheight\global\advance\vsize by -2\ht\partialpage
         \global\advance\vsize by 2\bigskipamount
         \global\advance\vsize by 1 pc}\eject % a little extra room; 1pc
     \output={\doublecolumnout\global\vsize=2\pageheight}
         \global\pagetextwidth=\columnwidth \global\hsize=\columnwidth}
% keeps footnotes on correct page
%
  \def\enddoublecolumns{\output={\balancecolumns\global\hsize=\pagewidth
                       \global\pagetextwidth=\pagewidth
                       \global\vsize=\pageheight
                       \unvbox255 }\eject\endgroup}
  \def\doublecolumnout{\splittopskip=\topskip \splitmaxdepth=\maxdepth
     \dimen@=\pageheight\advance\dimen@ by -\ht\partialpage
     \setbox0=\vsplit255 to\dimen@ \setbox2=\vsplit255 to \dimen@
     \onepageout\pagesofar \unvbox255 \penalty\outputpenalty}
  \def\pagesofar{\unvbox\partialpage
     \wd0=\columnwidth \wd2=\columnwidth \hbox to \pagewidth{\box0\hfil
     \columnrule \hfil \box2}}
  \def\columnrule{\vrule width \columnrulewidth height \ht2}
  \def\balancecolumns{\setbox0=\vbox{\unvbox255}\dimen@=\ht0
     \advance\dimen@ by \topskip \advance\dimen@ by-\baselineskip
     \advance\dimen@ by -2\ht\partialpage  % what if we begin and end on the
     \divide\dimen@ by2                    % same page?!?
     \splittopskip=\topskip
     {\vbadness=10000 \loop \global\setbox3=\copy0
        \global\setbox1=\vsplit3 to \dimen@
        \ifdim\ht3>\dimen@ \global\advance\dimen@ by1pt \repeat}
     \setbox0=\vbox to \dimen@{\unvbox1} \setbox2=\vbox to \dimen@{\dimen2=\dp3
     \unvbox3 \kern-\dimen2 \vfil }
     \pagesofar }
   \def\onepageout##1{ \setbox0=\vbox{##1} \dimen@=\dp0
     \shipout\vbox{ % here we define one page of output
     \makeheadline
     \vbox to \pageheight{
       \boxmaxdepth=\maxdepth
       \ifvoid\margin\else % marginal info is present
       \rlap{\kern31pc\vbox to 0pt{\kern4pt\box\margin\vss}}\fi
       \ifvoid\topins\else\unvbox\topins\vskip\skip\topins\fi
       ##1                                  % now insert the main information
       \vskip\pagebottomfiller
       \ifvoid\footins\else\vskip\skip\footins\footrule\unvbox\footins\fi
       \ifr@ggedbottom\kern-\dimen@ \vfil\fi}  %need a replacement for here
       \makefootline}
     \advancepageno\frontpagefalse}
\def\makeheadline{\vbox to\z@{\vskip-22.5\p@
  \hbox to \pagewidth{\vbox to8.5\p@{}\the\headline}\vss}\nointerlineskip}
   \def\makefootline{\baselineskip = 1.5\normalbaselineskip
             \hbox to \pagewidth{\the\footline}}
   \def\footrule{\dimen@=\prevdepth\nointerlineskip
      \vbox to 0pt{\vskip -0.25\baselineskip \hrule width 0.62\pagewidth \vss}
      \prevdepth=\dimen@ }
   \def\Vfootnote##1{\insert\footins\bgroup
      \interlinepenalty=\interfootnotelinepenalty \floatingpenalty=20000
      \singl@true\doubl@false\Tenpoint \hsize=\pagewidth
      \splittopskip=\ht\strutbox \boxmaxdepth=\dp\strutbox
      \leftskip=\footindent \rightskip=\z@skip
      \parindent=0.5\footindent \parfillskip=0pt plus 1fil
      \spaceskip=\z@skip \xspaceskip=\z@skip \footnotespecial
      \Textindent{##1}\footstrut\futurelet\next\fo@t}
% make the footnotes all have the correct size and the same footrule!
%% Make column separators for large one-column equations %%%%%%%%%%%%%%%%%%%
 \def\sp@cecheck##1{\dimen@=\pagegoal\advance\dimen@ by -\pagetotal
      \ifdim\dimen@<##1 \ifdim\dimen@>0pt \vfil\break \fi\fi}
 \def\endleftcolumn{\dimen@=\pagegoal\advance\dimen@ by -\pagetotal
      \ifdim\dimen@<\chapterminspace \ifdim\dimen@>0pt \vfil\break \fi
      \hbox{\vbox{\hrule width \columnwidth}\hbox to 0.4pt
      {\vrule height 10pt depth 0pt}\hfil}\fi}
 \def\beginrightcolumn{\dimen@=\pagegoal\advance\dimen@ by -\pagetotal
      \ifdim\dimen@<\chapterminspace \ifdim\dimen@>0pt \vfil\break \fi
      \hbox to \hsize{\hss\hbox{\vrule height 0pt depth 10pt
      \vbox{\hrule width \columnwidth}}}\fi}
}
%
%
%%%%%%%%%%%%%%%%%%%%%%%%%%%%%%%%%%%%%%%%%%%%%%%%%%%%%%%%%%%%%%%%%%%%%%%%
%
%   Now start the draftmode and preprintmode enhancement features
%                      (Homage to harvmac.tex)
%   Report any bugs to T. J. Allen
%       (tjallen@suhep.phy.syr.edu, tja@theory3.caltech.edu and
%        tjallen@wishep.physics.wisc.edu)
%
%%%%%%%%%%%%%%%%%%%%%%%%%%%%%%%%%%%%%%%%%%%%%%%%%%%%%%%%%%%%%%%%%%%%%%%%
%%
%%   Next, I define output routines, footnotes & related stuff.
%%   (The headline has been modified for draftmode and preprints
%%   may be produced in landscape form, two columns sideways)
%%
%%%%%%%%%%%%%%%%%%%%%%%%%%%%%%%%%%%%%%%%%%%%%%%%%%%%%%%%%%%%%%%%%%%%%%
%
%
\newif\ifpr@printstyle \pr@printstylefalse
\newbox\leftpage
\newdimen\fullhsize
\newdimen\titlepagewidth
\newdimen\pagetextwidth
\def\preprintstyle{%
       \message{(This will be printed PREPRINTSTYLE)}
       \let\lr=L
       \frontpagetrue
       \pr@printstyletrue
       \vsize=7truein
       \pagetextwidth=4.75truein
       \fullhsize=10truein
       \titlepagewidth=8truein
       \normalspace
       \Tenpoint
       \voffset=-.31truein
       \hoffset=-.46truein
       \iffrontpage\hsize=\titlepagewidth\else\hsize=\pagetextwidth\fi
 \output={%
    \iffrontpage
      \shipout\vbox{\special{\printertype}\makeheadline
      \hbox to \fullhsize{\hfill\pagebody\hfill}}
      \advancepageno
    \else
       \almostshipout{\leftline{\vbox{\pagebody\makefootline}}}\advancepageno
    \fi}
        \def\almostshipout##1{\if L\lr \count2=1
             \message{[\the\count0.\the\count1.\the\count2]}
        \global\setbox\leftpage=##1 \global\let\lr=R
                             \else \count2=2
        \shipout\vbox{\special{\printertype}
        \hbox to\fullhsize{\hfill\box\leftpage\hskip0.5truein##1\hfill}}
        \global\let\lr=L     \fi}
   \multiply\chapterminspace by 7 \divide\chapterminspace by 9
   \multiply\sectionminspace by 7 \divide\sectionminspace by 9
   \multiply\referenceminspace by 7 \divide\referenceminspace  by 9
   \multiply\chapterskip by 7 \divide\chapterskip  by 9
   \multiply\sectionskip  by 7 \divide\sectionskip  by 9
   \multiply\headskip   by 7 \divide\headskip by 9
   \multiply\baselineskip   by 7 \divide\baselineskip by 9
   \multiply\abovedisplayskip by 7 \divide\abovedisplayskip by 9
   \belowdisplayskip = \abovedisplayskip
\def\advancepageno{\if L\lr \gl@bal\advance\pagen@ by 1\fi
   \ifnum\pagenumber<0 \gl@bal\advance\pagenumber by -1
    \else\gl@bal\advance\pagenumber by 1 \fi
    \gl@bal\frontpagefalse  \swing@
    \gl@bal\hsize=\pagetextwidth}
} % end of preprintstyle specs

\tolerance=1000
\def\printertype{ps: }
%
% Default values for the fullsize document page

\paperheadline={\ifdr@ftmode\hfil\draftdate\else\hfill\fi}
\def\advancepageno{\gl@bal\advance\pagen@ by 1
   \ifnum\pagenumber<0 \gl@bal\advance\pagenumber by -1
    \else\gl@bal\advance\pagenumber by 1 \fi
    \gl@bal\frontpagefalse  \swing@
    \gl@bal\hsize=\pagetextwidth} %MODIFICATION
\def\papersize{\fullhsize=6.5in
               \pagetextwidth=6.5in
               \hsize=\fullhsize
               \vsize=9truein
               \hoffset=0.05 truein
               \voffset=-0.1truein
               \advance\hoffset by\HOFFSET
               \advance\voffset by\VOFFSET
               \pagebottomfiller=0pc
               \skip\footins=\bigskipamount
               \normalspace }
\papers
\def\lettersize{\fullhsize=6.5in
                \pagetextwidth=6.5in
                \hsize=\fullhsize
                \vsize=8.5in
                \hoffset=0in
                \voffset=0.5in
                \advance\hoffset by\HOFFSET
                \advance\voffset by\VOFFSET
                \pagebottomfiller=\letterbottomskip
                \skip\footins=\smallskipamount
                \multiply\skip\footins by 3
                \singlespace }
\def\semi{;\hfil\break}
%%%%%%%%%%%%%%%%%%%%%%%%%%%%%%%%%%%%%%%%%%%%%%%%%%%%%%%%%%%%%%%%%%%%%%%%
%%
%%   Here come chapter, section, subsection & appendix macros.
%%
%%%%%%%%%%%%%%%%%%%%%%%%%%%%%%%%%%%%%%%%%%%%%%%%%%%%%%%%%%%%%%%%%%%%%%%%
%
%  The following allows a shortcut for making titles bold etc.
%  Just use \chapterheadstyle={\bf} in the beginning of the
%  TeX file.
%
\newtoks\chapterheadstyle  \chapterheadstyle={\relax}
\def\chapter#1{{\the\chapterheadstyle\par \penalty-300 \vskip\chapterskip
   \spacecheck\chapterminspace
   \chapterreset \titlestyle{\chapterlabel.~#1}
   \nobreak\vskip\headskip \penalty 30000
   \message{(\the\chapternumber. #1)}
  {\pr@tect\wlog{\string\chapter\space \chapterlabel}} }}

\def\APPENDIX#1#2{{\the\chapterheadstyle\par\penalty-300\vskip\chapterskip
   \spacecheck\chapterminspace \chapterreset \xdef\chapterlabel{#1}
   \titlestyle{APPENDIX #2} \nobreak\vskip\headskip \penalty 30000
   \wlog{\string\Appendix~\chapterlabel} }}
\def\chapterreset{\gl@bal\advance\chapternumber by 1
   \ifnum\equanumber<0 \else\gl@bal\equanumber=0\fi
   \gl@bal\sectionnumber=0 \let\sectionlabel=\rel@x
   {\pr@tect
       \xdef\chapterlabel{{\the\chapterstyle{\the\chapternumber}}}}}%

%
%%%%%%%%%%%%%%%%%%%%%%%%%%%%%%%%%%%%%%%%%%%%%%%%%%%%%%%%%%%%%%%%%%%%%%%
%%
%%       Here is the draftmode feature
%%
%%       Use the following on the preliminary draft,
%%       puts time/date on each page in writes labels in margins
%%       and puts reference labels on the reference page.
%%       Putting \draft in the beginning of the paper causes it
%%       to be printed in draftmode.  use \nodraftlabels to get rid of
%%       eqn, ref, and fig labels in draft mode
%%
%%       Timestamp routine bug fixed October 30, 1991 by T.J. Allen
%%
%%%%%%%%%%%%%%%%%%%%%%%%%%%%%%%%%%%%%%%%%%%%%%%%%%%%%%%%%%%%%%%%%%%%%%%
%
\newif\ifdr@ftmode
\newtoks\r@flabeltoks
\def\draftmode{
   \pagetextwidth=6truein
   \fullhsize=6truein
   \titlepagewidth=6truein
   \normalspace
   \hoffset=0.3truein
   \voffset=0.2truein
   \hsize=\pagetextwidth
   \tenpoint
   \baselineskip=24pt plus 2pt minus 2pt
   \dr@ftmodetrue
   \message{ DRAFTMODE }
   \writedraftlabels
   \def\timestring{\begingroup
     \count0 = \time \divide\count0 by 60
     \count2 = \count0  % the hour
     \count4 = \time \multiply\count0 by 60
     \advance\count4 by -\count0   % the minute
     \ifnum\count4<10 \toks1={0} % get a leading zero.
     \else \toks1 = {}
     \fi
     \ifnum\count2<12 \toks0={a.m.} %
          \ifnum\count2<1 \count2=12 \fi% Make midnight `12'
     \else            \toks0={p.m.} %
           \ifnum\count2=12 % keep noon `12'
           \else
           \advance\count2 by -12 % keep afternoon times < 12
           \fi
     \fi
%    \ifnum\count2=0 \count2 = 12\fi % make midnight `12'. %  There seems to
%    be a bug in TeX when checking a count which has the value 0.
     \number\count2:\the\toks1 \number\count4\thinspace \the\toks0
   \endgroup}%
   \def\draftdate{{{\tt preliminary version:}\space{\rm
                                  \timestring\quad\the\date}}}
\def\R@FWRITE##1{\ifreferenceopen \else \gl@bal\referenceopentrue
     \immediate\openout\referencewrite=\jobname.refs
     \toks@={\begingroup \refoutspecials \catcode`\^^M=10 }%
     \immediate\write\referencewrite{\the\toks@}\fi
     \immediate\write\referencewrite%
     {\noexpand\refitem{\the\r@flabeltoks[\the\referencecount]}}%
     \p@rse@ndwrite \referencewrite ##1}
\def\refitem##1{\r@fitem{##1}}
\def\REF##1##2{\reflabel##1 \REFNUM ##1\REFWRITE{\ignorespaces ##2}}
\def\Ref##1##2{\reflabel##1 \Refnum ##1\REFWRITE{ ##2}}
\def\REFS##1##2{\reflabel##1 \REFNUM ##1%
\gl@bal\lastrefsbegincount=\referencecount\REFWRITE{ ##2}}
\def\refs{\REFS\?}
\def\refc{\REF\?}
\let\refscon=\refc       \let\REFSCON=\REF
}
\def\nodraftlabels{\def\leqlabel##1{}\def\eqlabel##1{}\def\reflabel##1{}%
\def\leqlabel##1{}}
\def\writedraftlabels{
  \def\eqlabel##1{{\escapechar-1\rlap{\sevenrm\hskip.05in\string##1}}}%
  \def\leqlabel##1{{\escapechar-1\llap{\sevenrm\string##1\hskip.05in}}}%
  \def\reflabel##1{\r@flabeltoks={{\escapechar-1\sevenrm\string##1\hskip.06in%
}}}}

\nodraftlabels   % Make the default mode no labels
\dr@ftmodefalse  % Turn off draftmode
%
%%%%%%%%%%%%%%%%%%%%%%%%%%%%%%%%%%%%%%%%%%%%%%%%%%%%%%%%%%%%%%%%%%%%%%%%
%%
%%   Here come macros for equation numbering.
%%   (Equation numbers are modified in draft mode)
%%
%%   Sections are automatically numbered independently, unless
%%   one puts the command \sequentialequations
%%
%%%%%%%%%%%%%%%%%%%%%%%%%%%%%%%%%%%%%%%%%%%%%%%%%%%%%%%%%%%%%%%%%%%%%%%%
%
%

\def\eqn#1{\eqno\eqname{#1}\eqlabel#1}
 %MODIFICATION
%
%
\def\eqinsert#1{\noalign{\dimen@=\prevdepth \nointerlineskip
   \setbox0=\hbox to\displaywidth{\hfil #1}
   \vbox to 0pt{\kern 0.5\baselineskip\hbox{$\!\box0\!$}\vss}
   \prevdepth=\dimen@}}  %MODIFICATION

%

 %MODIFICATION

%
 %MODIFICATION
%
%\def\sequentialequations{\equanumber=-1}
%\def\sequentialequations{\rel@x \if\equanumber<0 \else
%  \gl@bal\equanumber=-\equanumber \gl@bal\advance\equanumber by -1 \fi }
%                         %MODIFICATION
%
%%%%%%%%%%%%%%%%%%%%%%%%%%%%%%%%%%%%%%%%%%%%%%%%%%%%%%%%%%%%%%%%%%%%%%%%
%%
%%  Here come reference macros  (Modified for the new version of phyzzx)
%%
%%%%%%%%%%%%%%%%%%%%%%%%%%%%%%%%%%%%%%%%%%%%%%%%%%%%%%%%%%%%%%%%%%%%%%%%
%
%
\def\refout{\par\penalty-400\vskip\chapterskip
   \spacecheck\referenceminspace
   \ifreferenceopen \Closeout\referencewrite \referenceopenfalse \fi
   \line{\ifpr@printstyle\twelverm\else\fourteenrm\fi
         \hfil REFERENCES\hfil}\vskip\headskip
   \input \jobname.refs
   }
\def\ACK{\par\penalty-100\medskip \spacecheck\sectionminspace
   \line{\ifpr@printstyle\twelverm\else\fourteenrm\fi
      \hfil ACKNOWLEDGEMENTS\hfil}\nobreak\vskip\headskip }
\def\tabout{\par\penalty-400
   \vskip\chapterskip\spacecheck\referenceminspace
   \iftableopen \Closeout\tablewrite \tableopenfalse \fi
   \line{\ifpr@printstyle\twelverm\else\fourteenrm\fi\hfil TABLE CAPTIONS\hfil}
   \vskip\headskip
   \input \jobname.tabs
   }
\def\figout{\par\penalty-400
   \vskip\chapterskip\spacecheck\referenceminspace
   \iffigureopen \Closeout\figurewrite \figureopenfalse \fi
   \line{\ifpr@printstyle\twelverm\else\fourteenrm\fi\hfil FIGURE
CAPTIONS\hfil}
   \vskip\headskip
   \input \jobname.figs
   }
\def\masterreset{\begingroup\hsize=\pagetextwidth
   \global\pagenumber=1 \global\chapternumber=0
   \global\equanumber=0 \global\sectionnumber=0
   \global\referencecount=0 \global\figurecount=0 \global\tablecount=0
   \endgroup}
%

% % % % % % % % % % % % % % % % % % % % % % % % % % % % % % % % % % % %
%%%%%%%%%%%%%%%%%%%%%%%%%%%%%%%%%%%%%%%%%%%%%%%%%%%%%%%%%%%%%%%%%%%%%%%%%
%%
%%      Various little user definitions
%%
%%%%%%%%%%%%%%%%%%%%%%%%%%%%%%%%%%%%%%%%%%%%%%%%%%%%%%%%%%%%%%%%%%%%%%%%%%%%%%
%

\def\12{{1\over2}}
\def\etal{{\it et al.\ }}

\def\sla{\raise.15ex\hbox{$/$}\kern-.57em}
\def\leaderfill{\leaders\hbox to 1em{\hss.\hss}\hfill}
\def\dual{{\,^*\kern-.20em}}
\def\bx{{\vcenter{\hrule height 0.4pt
      \hbox{\vrule width 0.4pt height 10pt \kern 10pt
        \vrule width 0.4pt}
      \hrule height 0.4pt}}}
\def\inner{\,{\vcenter{
      \hbox{ \kern 4pt
        \vrule width 0.5pt height 7pt}
      \hrule height 0.5pt}}\,}
\def\sqr#1#2{{\vcenter{\hrule height.#2pt
      \hbox{\vrule width.#2pt height#1pt \kern#1pt
        \vrule width.#2pt}
      \hrule height.#2pt}}}
\def\rect#1#2#3#4{{\vcenter{\hrule height#3pt
      \hbox{\vrule width#4pt height#1pt \kern#1pt
        \vrule width#4pt}
      \hrule height#3pt}}}

\def\bx{{\vcenter{\hrule height 0.4pt
      \hbox{\vrule width 0.4pt height 10pt \kern 10pt
        \vrule width 0.4pt}
      \hrule height 0.4pt}}}

\def\up#1{\leavevmode \raise.16ex\hbox{#1}}
\def\twiddle{\lower.9ex\rlap{$\kern-.1em\scriptstyle\sim$}}
\def\bigtwiddle{\lower1.ex\rlap{$\sim$}}
\def\gtwid{\mathrel{\raise.3ex\hbox{$>$\kern-.75em\lower1ex\hbox{$\sim$}}}}
\def\ltwid{\mathrel{\raise.3ex\hbox{$<$\kern-.75em\lower1ex\hbox{$\sim$}}}}
\def\square{\kern1pt\vbox{\hrule height 1.2pt\hbox{\vrule width 1.2pt\hskip 3pt
   \vbox{\vskip 6pt}\hskip 3pt\vrule width 0.6pt}\hrule height 0.6pt}\kern1pt}
\def\tdot#1{\mathord{\mathop{#1}\limits^{\kern2pt\ldots}}}

\def\pmb#1{\setbox0=\hbox{#1}    %  POOR MAN'S BOLD
  \kern-.025em\copy0\kern-\wd0
  \kern  .05em\copy0\kern-\wd0
  \kern-.025em\raise.0433em\box0 }

                           %%DALEMBERTIAN, USED TO BE \box

\hyphenation{anom-aly anom-alies coun-ter-term coun-ter-terms}
\def\inv{^{\raise.15ex\hbox{${\scriptscriptstyle -}$}\kern-.05em 1}}

\def\Dsl{\,\raise.15ex\hbox{/}\mkern-13.5mu D} %this one can be subscripted
\def\dsl{\raise.15ex\hbox{/}\kern-.57em\partial}

       %pound sterling
\def\boxeqn#1{\vcenter{\vbox{\hrule\hbox{\vrule\kern3pt\vbox{\kern3pt
        \hbox{${\displaystyle #1}$}\kern3pt}\kern3pt\vrule}\hrule}}}
\def\mbox#1#2{\vcenter{\hrule \hbox{\vrule height#2in
                \kern#1in \vrule} \hrule}}  %e.g. \mbox{.1}{.1}
%
%%       matters of taste
%%  \def\tilde{\widetilde} \def\bar{\overline} \def\hat{\widehat}
%%

\def\darr#1{\raise1.5ex\hbox{$\leftrightarrow$}\mkern-16.5mu #1}
 %pound sterling
\def\roughly#1{\raise.3ex\hbox{$#1$\kern-.75em\lower1ex\hbox{$\sim$}}}
%
%
%%%%%%%%%%%%%%%%%%%%%%%%%%%%%%%%%%%%%%%%%%%%%%%%%%%%%%%%%%%%%%%%%%%%%%%%
%%
%%   Miscellaneous macros
%%
%%%%%%%%%%%%%%%%%%%%%%%%%%%%%%%%%%%%%%%%%%%%%%%%%%%%%%%%%%%%%%%%%%%%%%%%
%
\def\ack{\ACK}   % make new phyzzx compatible with old phyzzx
\font\titlerm=cmr10 scaled \magstep 4
\def\TITLEPAGE{\frontpagetrue\pageno=1\pagenumber=1}
%
%%%%%%%%%%%%%%%%%%%%%%%%%%%%%%%%%%%%%%%%%%%%%%%%%%%%%%%%%%%%%%%%%%%%%%%%%%%
%%%%%%%%%%%%%%%%%%%  Caltech Preprint Pub Block   %%%%%%%%%%%%%%%%%%%%%%%%%
%%%%%%%%%%%%%%%%%%%%%%%%%%%%%%%%%%%%%%%%%%%%%%%%%%%%%%%%%%%%%%%%%%%%%%%%%%%
\def\CALT#1{\hbox to\hsize{\tenpoint \baselineskip=12pt
        \hfil\vtop{\hbox{\strut CALT-68-#1}
        \hbox{\strut DOE RESEARCH AND}
        \hbox{\strut DEVELOPMENT REPORT}}}}

%%%%%%%%%%%%%%%%%%%%%%%%%%%%%%%%%%%%%%%%%%%%%%%%%%%%%%%%%%%%%%%%%%%%%%%%%%%
%%%%%%%%%%%%%%%%%%%%%%  SU Preprint Pub Block    %%%%%%%%%%%%%%%%%%%%%%%%%%
%%%%%%%%%%%%%%%%%%%%%%%%%%%%%%%%%%%%%%%%%%%%%%%%%%%%%%%%%%%%%%%%%%%%%%%%%%%

%PUT IN PREPRINT # AND THE DATE = #2

%

\def\WISCONSIN{\vskip15pt\vbox{\hbox{\centerline{\it Department of Physics}}
        \vskip 0pt
  \hbox{\centerline{\it 1150 University Avenue}}\vskip 0pt
  \hbox{\centerline{\it University of Wisconsin, Madison, WI 53706 USA}}}}

\def\TITLE#1{\vskip 1in \centerline{\titlerm #1}}
\def\MORETITLE#1{\vskip 19pt \centerline{\titlerm #1}}
\def\AUTHOR#1{\vskip .5in \centerline{#1}}

\def\ABSTRACT#1{\vskip .5in \vfil \centerline{\twelvepoint \bf Abstract}
        #1 \vfil}
\def\ENDTITLEPAGE{\vfill\eject\pageno=2\pagenumber=2}%\hsize=\pagetextwidth}

\def\underwig#1{{
\setbox0=\hbox{$#1$}
\setbox1=\hbox{}
\wd1=\wd0
\ht1=\ht0
\dp1=\dp0
\setbox2=\hbox{$\rm\widetilde{\box1}$}
\dimen@=\ht2 \advance \dimen@ by \dp2 \advance \dimen@ by 1.5pt
\ht2=0pt \dp2=0pt
\hbox to 0pt{$#1$\hss} \lower\dimen@\box2
}}
\def\bunderwig#1{{
\setbox0=\hbox{$#1$}
\setbox1=\hbox{}
\wd1=\wd0
\ht1=\ht0
\dp1=\dp0
\setbox2=\hbox{$\seventeenrm\widetilde{\box1}$}
\dimen@=\the\ht2 \advance \dimen@ by \the\dp2 \advance \dimen@ by 1.5pt
\ht2=0pt \dp2=0pt
\hbox to 0pt{$#1$\hss} \lower\dimen@\box2
}}
\def\journal#1&#2(#3){\unskip, \sl #1~\bf #2 \rm (19#3) }
                    % Journal reference. Alignment
                    % tabs & set off name, vol, year, page
\def\npjournal#1&#2&#3&#4&{\unskip, #1~\rm #2 \rm (#3) #4}
\gdef\prjournal#1&#2&#3&#4&{\unskip, #1~\bf #2, \rm #4 (#3)}

\def\ket#1{\left| #1\right\rangle}
\def\VEV#1{\left\langle #1\right\rangle}

\let\int=\intop         
\catcode`@=12 % at signs are no longer letters
\masterreset
%%%%%%%%%%%%%%%%%%%% End of phyzzx.local %%%%%%%%%%%%%%%%%%%%%%%%%%%%%%%%
%%%%%%%%%%%%%%%%%%%%%%% Start of Paper %%%%%%%%%%%%%%%%%%%%%%%%%%%%%%%%%%
%use the phyzzx macropackage
%\baselineskip=13pt plus 0.2pt minus 0.1pt
%\draft
%\preprintstyle
\tenpoint
\normalspace
%\doublespace
\chapterheadstyle={\bf}
\def\Re{{\rm Re\,\,}}
\def\Im{{\rm Im\,\,}}
\overfullrule=0pt
\TITLEPAGE
\rightline{{\tenpoint\baselineskip=12pt
           \vtop{\hbox{\strut MAD/TH-92-04}
                 \hbox{\strut hep-th/9302137}
                 \hbox{\strut December 1992 }}}}
\TITLE{ Harmonic BRST Quantization of Systems with Irreducible }
\MORETITLE{ Holomorphic Boson and Fermion Constraints }
\AUTHOR{ Theodore J. Allen {\it and} Dennis B. Crossley }
\WISCONSIN

\ABSTRACT{ We show that the harmonic Becchi-Rouet-Stora-Tyutin method of
quantizing bosonic systems with second-class constraints or first-class
holomorphic constraints extends to systems having both bosonic and fermionic
second-class or first-class holomorphic constraints. Using a limit  argument,
we show that the harmonic BRST modified path integral reproduces the correct
Senjanovic measure. }

\ENDTITLEPAGE

\chapter{ Introduction }

A new implementation of Becchi-Rouet-Stora-Tyutin quantization was introduced
recently by one of the authors in order to quantize theories with bosonic
holomorphic constraints.\Ref\HBRST{T.J.~Allen, {\it Phys. Rev. } {\bf D43}
(1991) 3442.}  The main assumption of the method is that there is an algebra of
first-class constraints, some of which are not real-valued, but rather are
holomorphic, functions and that some subset of the holomorphic constraints,
together with their complex conjugates, are second-class.  That is, the matrix
of Poisson brackets of the holomorphic with their anti-holomorphic partners is
not weakly vanishing.  If the matrix of Poisson brackets were weakly vanishing,
one could simply take the real and imaginary parts of the constraints as being
separate first-class constraints and ignore altogether the difficulties of any
holomorphic structure.  Keeping the holomorphic structure is impossible in a
standard  Becchi-Rouet-Stora-Tyutin---Batalin-Fradkin-Vilkovisky quantization.
The quantum constraints and, hence, the BRST-BFV charge operator will not be
hermitian, making it impossible to decouple the unphysical states from the
physical ones.

The number of systems to which our method applies is in principle quite large.
Kalau\Ref\Kalau{W.~Kalau, {\it Int. J. Mod. Phys.} {\bf A8} (1993) 391.} has
shown that any set of an even number of second-class constraints may be split
into holomorphic and anti-holomorphic algebras, but has noted that the split
may not be computationally useful, either because the quantum algebra may have
anomalies or because the holomorphic constraints are computationally intricate.
Perhaps the most important example in which a holomorphic structure is useful
is that of Ashtekar's new canonical variables reformulation of general
relativity,\Ref\Ashtekar{A.~Ashtekar, {\sl Lectures on Non-perturbative
Canonical Gravity,} Lecture notes prepared in collaboration with R.S.~Tate.
(World Scientific Books, Singapore, 1991)\semi\^^M C.~Rovelli, {\it Class.
Quantum Grav.} {\bf 8} (1991) 1613.} where all the constraints are first-class
holomorphic and polynomial when written in terms of the  self-dual spin
connection.
%%%%%%%%%%%%%%%%%%%%%%%%%%%%%%%%%%%%%%%%%%%%%%%%%%%%%%%%%%%%%%%%%%%%%%%%%%%%%
\REF\NPS{E.~Nissimov, S.~Pacheva, and S.~Solomon, {\it Nucl. Phys.} {\bf B297}
(1988) 349.}\REF\tja{T.J.~Allen, Caltech Ph.D. Thesis
(1988).}\REF\Aoyama{S.~Aoyama, J.~Kowalski-Glikman, J.~Lukierski and  J.W.~van
Holten, {\it Phys. Lett. }{\bf 216B} (1989) 133.}
%%%%%%%%%%%%%%%%%%%%%%%%%%%%%%%%%%%%%%%%%%%%%%%%%%%%%%%%%%%%%%%%%%%%%%%%%%%%%
Other examples are the $D = 10$ harmonic superstring and
superparticle,\refmark{\NPS,\tja}  the Brink-Schwarz superparticle in four
dimensions\refmark{\tja,\Aoyama} and certain coadjoint orbit theories such as
particle spin dynamics on a Lie group.\Ref\SpinLag{A.P. Balachandran,  S.
Borchardt and A. Stern, {\it Phys. Rev. }{\bf D17} (1978) 3247\semi\^^M A.P.
Balachandran, G. Marmo, B.-S. Skagerstam and A. Stern, {\sl Gauge Symmetries
and Fibre Bundles,} Springer-Verlag Lecture Notes in Physics 188,
(Springer-Verlag, Berlin and Heidelberg, 1983)\semi\^^M A.P. Balachandran, M.
Bourdeau, and S. Jo, {\it Int. J. Mod. Phys. }{\bf A5} (1990) 2423\semi\^^M E.
Gates, R. Potting, C. Taylor, and B. Velikson, {\it Phys. Rev. Lett.} {\bf 63}
(1989) 2617.}
%%%%%%%%%%%%%%%%%%%%%%%%%%%%%%%%%%%%%%%%%%%%%%%%%%%%%%%%%%%%%%%%%%%%%%%%%%%%%
\REF\Kowalski{J.~Kowalski-Glikman, {\it Phys. Lett.} {\bf B245} (1990) 79.
{\it ``On Gupta-Bleuler Quantization of Hamiltonian Systems with Anomalies,''}
hep-th preprint 9211028, to appear in {\it Ann. Phys.}}
%%%%%%%%%%%%%%%%%%%%%%%%%%%%%%%%%%%%%%%%%%%%%%%%%%%%%%%%%%%%%%%%%%%%%%%%%%%%%
We remark that this splitting can also be used with the operatorial
quantization in the case that classical first-class constraints become
anomalous in the quantum theory, but our path-integral may need to be modified
along the lines of ref.\ [\Kowalski]. Operatorial constructions somewhat
different from ours have been given  by Hasiewicz \etal\Ref\Hasiewicz{Z.
Hasiewicz, J. Kowalski-Glikman, J. Lukierski, and J.W. van Holten, {\it Phys.
Lett. }{\bf 217B} (1989) 95.} and recently by Marnelius.\Ref\Marnelius{R.
Marnelius, {\it Nucl.\ Phys.\ } {\bf B 370} (1992) 165.}  The only example of a
system known to us to which our method does not  apply in principle is Berezin
and Marinov's\Ref\Fermion{F. A. Berezin and  M. S. Marinov, {\it Ann. Phys.\/}
(N.Y.) {\bf 104} (1977) 336\semi F. Bordi and R. Casalbuoni, {\it Phys.
Lett.\/} {\bf B 93} (1980) 308\semi T. J. Allen, {\it Phys. Lett.\/} {\bf B
214} (1988) 87.} action for a massive point fermion which has an odd number of
second-class fermionic  constraints.

The present work extends the harmonic BRST-BFV quantization scheme to the case
of both bosonic and fermionic constraints.  In section 2 we review the use of
the harmonic BRST-BFV method for systems with bosonic constraints. In section 3
we demonstrate the extension of this method to systems with fermionic
constraints by applying it to the most trivial case, that of a single fermionic
constraint.  In section 4 we treat the general case of an arbitrary number of
bosonic and fermionic constraints.  We demonstrate that in a certain limit, the
modification of the  path integral reproduces the correct
Senjanovic\Ref\Senjanovic{P.  Senjanovic, {\it Ann. Phys.} (N.Y.) {\bf 100}
(1976) 227.} measure for  second-class constraints.  Throughout the analysis we
have assumed that the constraints are irreducible.

\chapter{ The Harmonic BRST-BFV Method for Bosonic Constraints }

The main assumption that we start from is the existence of an algebra of
holomorphic constraints, $\cal A$, closed under Poisson brackets, whose matrix
of Poisson brackets with the complex conjugate algebra, $\bar{\cal A}$, has
non-vanishing determinant, even weakly,
%%%%%%%%%%%%%%%%%%%%%%%%%%%%%%%%%%%%%%%%%%%%%%%%%%%%%%%%%%%%%%%%%%%%%%%%%%%%%
$$ \det_{(ij)}  \{ a_i, \bar{a}_j\} \not\approx 0, \quad a_i\in{\cal A},\quad
\bar{a}_j \in\bar{\cal A}.\eqn\assume$$
%%%%%%%%%%%%%%%%%%%%%%%%%%%%%%%%%%%%%%%%%%%%%%%%%%%%%%%%%%%%%%%%%%%%%%%%%%%%%

If the determinant in \assume\ were zero, we could find a subalgebra, ${\cal
B}\subset{\cal A}$, such that the determinant was non-vanishing for
$a_i\in{\cal B}$, $\bar{a}_j\in\bar{\cal B}$.  We could then take the real and
imaginary parts of the remaining constraints $a_i\in{\cal A}\\ {\cal B}$ and
treat them separately.

We also assume that there may be some first-class constraints, $\cal F$, as
well, which we may take to be real.  To construct the necessary operators, we
look at the one-parameter set of algebras, ${\cal F}\oplus t{\cal A}$,
consisting of the first-class constraints and holomorphic constraints scaled by
an arbitrary real parameter $t$.  It is well known that there exists a
fermionic function, the  formal BRST charge, which has vanishing Poisson
bracket with itself,\Ref\Henneaux{M.~Henneaux, {\it Phys. Rep.\/} {\bf 126}
(1985) 1.}
%%%%%%%%%%%%%%%%%%%%%%%%%%%%%%%%%%%%%%%%%%%%%%%%%%%%%%%%%%%%%%%%%%%%%%%%%%%%%
$$\eqalign{Q({\cal F}\oplus t{\cal A})&=\eta_If_I+t\eta_ia_i+\ldots\cr
&=\Omega + t\Theta,\cr \{Q,Q\} &= 0.\cr}\eqn\Charges$$
%%%%%%%%%%%%%%%%%%%%%%%%%%%%%%%%%%%%%%%%%%%%%%%%%%%%%%%%%%%%%%%%%%%%%%%%%%%%%
In eq.\ \Charges, the $\eta$'s are ghost variables, which are anticommuting
when the constraints are bosonic.  It is worth pointing out that the charge
$\Omega$ is not necessarily identical to the BRST charge used when the
second-class constraints are implemented using Dirac brackets.  For an example
of this, see ref.\ [\HBRST].

Because the parameter $t$ is arbitrary, the last equality of \Charges\ implies
the following relations,
%%%%%%%%%%%%%%%%%%%%%%%%%%%%%%%%%%%%%%%%%%%%%%%%%%%%%%%%%%%%%%%%%%%%%%%%%%%%%
$$\{\Omega,\Omega\} = \{\Omega,\Theta\} = \{\Theta,\Theta\} =
0.\eqn\relations$$
%%%%%%%%%%%%%%%%%%%%%%%%%%%%%%%%%%%%%%%%%%%%%%%%%%%%%%%%%%%%%%%%%%%%%%%%%%%%%

Unfortunately, the charge $Q$ defined in \Charges\ is not real, so it is not
suitable to use its quantum version, $\hat Q$,  to define physical states.  It
is necessary for the existence of a BRST cohomology that the operator $\hat Q$
be either hermitian or anti-hermitian.  This is because the equivalence
relation $\ket{\Psi} \cong \ket{\Psi} + \hat{Q}\ket{\Phi}$, with $\ket{\Phi}$
arbitrary, must be compatible with the inner product on the Hilbert space. In
other words, states that are $\hat Q$-exact must be orthogonal to the physical
$\hat Q$-closed (and $\hat Q$-exact) states and, therefore, have zero norm.  To
have the  decoupling
%%%%%%%%%%%%%%%%%%%%%%%%%%%%%%%%%%%%%%%%%%%%%%%%%%%%%%%%%%%%%%%%%%%%%%%%%%%%%
$$\VEV{\Psi\Big|\hat{Q}\Phi}=\VEV{\hat{Q}^\dagger\Psi\Big|\Phi}=0, \eqn\QHerm$$
%%%%%%%%%%%%%%%%%%%%%%%%%%%%%%%%%%%%%%%%%%%%%%%%%%%%%%%%%%%%%%%%%%%%%%%%%%%%%
for all physical states $\ket{\Psi}$ and all arbitrary $\ket{\Phi}$, it is
necessary that $\hat Q = \pm\hat{Q}^\dagger$.

It is useful to use the operators $\hat\Omega$ and $\hat\Theta$ separately.
Physical states are defined to be those which are in the cohomology of
$\hat\Omega$ and are annihilated by both $\hat\Theta$ and its adjoint,
$\hat{\bar\Theta}$,
%%%%%%%%%%%%%%%%%%%%%%%%%%%%%%%%%%%%%%%%%%%%%%%%%%%%%%%%%%%%%%%%%%%%%%%%%%%%%
$$\eqalign{\hat\Theta\ket{{\rm phys}}&=\hat{\bar\Theta}\ket{{\rm
phys}}=\hat\Omega\ket{{\rm phys}} =0,\cr  \ket{{\rm phys}} &\cong \ket{{\rm
phys}} + \hat\Omega\ket{{\rm anything}}.\cr}\eqn\PhysStates$$
%%%%%%%%%%%%%%%%%%%%%%%%%%%%%%%%%%%%%%%%%%%%%%%%%%%%%%%%%%%%%%%%%%%%%%%%%%%%%
These states are harmonic in the sense that they are annihilated by the
Laplacian $\{\hat\Theta,\hat{\bar\Theta}\}$.

The ghosts used in $\Theta$ and $\bar\Theta$ are different from the usual BRST
ghosts.  There are two inequivalent complex structures one can impose on
fermionic phase space variables.  The one closest in analogy to the bosonic
oscillator has the canonically conjugate variables, $\xi$ and $\xi^*=
\bar{\xi}$,   which are also complex conjugates of one another.  The real and
imaginary parts of $\xi$, in addition to being real, are also canonically
self-conjugate,
%%%%%%%%%%%%%%%%%%%%%%%%%%%%%%%%%%%%%%%%%%%%%%%%%%%%%%%%%%%%%%%%%%%%%%%%%%%%%
$$\eqalign{\{\xi,\bar{\xi}\}&=-i,\quad \xi = {1\over\sqrt{2}}(\rho + i\pi),\cr
\{\rho,\rho\}&= \{\pi,\pi\} = -i,\quad  \{\rho,\pi\}=0.\cr}\eqn\Hghosts$$
%%%%%%%%%%%%%%%%%%%%%%%%%%%%%%%%%%%%%%%%%%%%%%%%%%%%%%%%%%%%%%%%%%%%%%%%%%%%%
\REF\HenneauxII{M.~Henneaux, {\sl Classical Foundations of BRST Symmetry},
(Bibliopolis, Naples, 1988)  pp.~45-50.}\REF\Govaerts{Jan~Govaerts, {\sl
Hamiltonian Quantization and Constrained Dynamics}, (Leuven University Press,
Leuven, 1991) p.\ 69.}  These are the ghosts that we will require.  They differ
from the $(b,c)$ ghosts\refmark{\HenneauxII,\Govaerts} of the usual BRST-BFV
formalism. The $(b,c)$ ghosts are real and are not canonically self-conjugate,
%%%%%%%%%%%%%%%%%%%%%%%%%%%%%%%%%%%%%%%%%%%%%%%%%%%%%%%%%%%%%%%%%%%%%%%%%%%%%
$$\{b,c\} = -i,\quad \{b,b\} = \{c,c\} = 0, \quad b^*=b,\quad c^*
=c.\eqn\bcghosts$$
%%%%%%%%%%%%%%%%%%%%%%%%%%%%%%%%%%%%%%%%%%%%%%%%%%%%%%%%%%%%%%%%%%%%%%%%%%%%%
The symmetry of the fermionic Poisson bracket allows the existence of these two
inequivalent complex structures on a pair of canonically conjugate fermionic
variables. Bosons have only one such structure.  The ghosts used to construct
$\Theta$  for the holomorphic constraints $a_i\in{\cal A}$ are of the first
type \Hghosts, while the ghosts used in $\Omega$ are the $(b,c)$ ghosts.

In ref.\ [\HBRST] a Hamiltonian path integral construction was given, analogous
to the BFV\Ref\BFV{ E. S. Fradkin and G. A. Vilkovisky, {\it Phys. Lett.} {\bf
55B} (1975) 224\semi\^^M I. A. Batalin and G. A. Vilkovisky, {\it Phys. Lett.}
{\bf 69B} (1977) 309\semi\^^M E. S. Fradkin and T. E. Fradkina, {\it Phys.
Lett.} {\bf 72B} (1978) 343.} construction explained in great detail in ref.\
[\Henneaux].  When the assumption of the holomorphic BRST-BFV method applies,
the Hamiltonian path integral is
%%%%%%%%%%%%%%%%%%%%%%%%%%%%%%%%%%%%%%%%%%%%%%%%%%%%%%%%%%%%%%%%%%%%%%%%%%%%%
$${\cal Z}_{\Psi,\beta}=\int{\cal D}\mu\,\exp\,{i\over\hbar} \int
dt(i\bar\xi\dot\xi + p\dot q + b\dot c + \bar c\dot{\bar b} + \pi\dot\lambda -
H_{BRST} - \beta\{\Theta,\bar\Theta\} - \{\Omega,\Psi\}).\eqn\HBRSTpath$$
%%%%%%%%%%%%%%%%%%%%%%%%%%%%%%%%%%%%%%%%%%%%%%%%%%%%%%%%%%%%%%%%%%%%%%%%%%%%%
If the fermionic gauge-fixing parameter $\Psi$ and the constant $\beta$ are
both imaginary, then the path integral is manifestly unitary.  One can show
that  the path integral \HBRSTpath\ is invariant under infinitesimal variations
of $\Psi$ and $\beta$.  The ``extra'' piece,
$\exp(-i\beta\{\Theta,\bar\Theta\})$, in \HBRSTpath\ in the limit of $\beta\to
\infty$ becomes the correct Senjanovic measure for
second-class constraints and eliminates the ghost degrees of freedom as well,
%%%%%%%%%%%%%%%%%%%%%%%%%%%%%%%%%%%%%%%%%%%%%%%%%%%%%%%%%%%%%%%%%%%%%%%%%%%%%
$$\lim_{\beta\to \infty} e^{-i\beta\{\Theta,\bar\Theta\}} =
(-1)^N\pi^N\delta^N(\xi_i)\delta^N(\bar\xi_j)\det( i\{a_i,\bar{a}_j\})
\delta^N({\rm Re}\, a_i)\delta^N({\rm Im}\, a_j).\eqn\deltafcnI$$
%%%%%%%%%%%%%%%%%%%%%%%%%%%%%%%%%%%%%%%%%%%%%%%%%%%%%%%%%%%%%%%%%%%%%%%%%%%%%

\chapter{ Fermionic Constraints }

We first consider the simple case of a single holomorphic fermionic constraint,
$\phi \approx 0$.  Since both the real and imaginary parts of $\phi$ must
weakly vanish, it follows that the complex conjugate constraint also vanishes,
$\bar\phi \approx 0$.  The harmonic BRST-BFV method then introduces a pair of
bosonic ghosts,  which are both complex conjugate and canonically conjugate.
A single fermionic constraint can have only a very simple algebra, although one
that is more general than that of a single bosonic constraint.  That algebra is
%%%%%%%%%%%%%%%%%%%%%%%%%%%%%%%%%%%%%%%%%%%%%%%%%%%%%%%%%%%%%%%%%%%%%%%%%%%%%
$$ \{\phi,\phi\} = \gamma\phi, \eqn\fermiA$$
%%%%%%%%%%%%%%%%%%%%%%%%%%%%%%%%%%%%%%%%%%%%%%%%%%%%%%%%%%%%%%%%%%%%%%%%%%%%%
where $\gamma$ is a fermionic function on phase space.  For simplicity, we
assume that the  brackets between $\phi$ and $\bar\phi$ are those of the
fermionic oscillator, $\{\phi,\bar\phi\} = -i$. We consider explicitly the case
in which $\gamma$ is a (Grassmann odd) constant. In this case the method yields
a state identical to that  of a single bosonic constraint, but with the roles
of the original and the ghost variables interchanged.  This is a consequence of
the $OSp(1,\!1\,|\,2)$ invariance of the system.\refmark{\HBRST}

The harmonic BRST charges for \fermiA\ are
%%%%%%%%%%%%%%%%%%%%%%%%%%%%%%%%%%%%%%%%%%%%%%%%%%%%%%%%%%%%%%%%%%%%%%%%%%%%%
$$\hat\Theta = \hat{c}\hat\phi + {i\over2}\hat\gamma\hat{\bar{c}}\hat{c}
\hat{c}, \quad \hat{\bar\Theta} = \hat{\bar{c}}\hat{ \bar\phi} -
{i\over2}\hat{\bar\gamma}\hat{\bar{c}}\hat{\bar{c}}\hat{c}.\eqn\chargesI$$
%%%%%%%%%%%%%%%%%%%%%%%%%%%%%%%%%%%%%%%%%%%%%%%%%%%%%%%%%%%%%%%%%%%%%%%%%%%%%
The most general state in the ghost-enlarged Hilbert space is  a sum of
products of ghost states, $\ket{n}_c$, of occupation number $n$  with states,
$\ket{\psi_n}$, of the original Hilbert space: $\ket{\Psi} =
\sum_{n=0}^{\infty} \ket{n}_c\ket{\psi_n}$.  The harmonicity conditions
\PhysStates\ yield the physical state
%%%%%%%%%%%%%%%%%%%%%%%%%%%%%%%%%%%%%%%%%%%%%%%%%%%%%%%%%%%%%%%%%%%%%%%%%%%%%
$$ \ket{{\rm phys}} = \ket{0}_c\ket{\psi_0}, \quad \hat{\bar\phi} \ket{\psi_0}
= 0.\eqn\stateI$$
%%%%%%%%%%%%%%%%%%%%%%%%%%%%%%%%%%%%%%%%%%%%%%%%%%%%%%%%%%%%%%%%%%%%%%%%%%%%%

The Poisson bracket of $\Theta$ and $\bar\Theta$ is
%%%%%%%%%%%%%%%%%%%%%%%%%%%%%%%%%%%%%%%%%%%%%%%%%%%%%%%%%%%%%%%%%%%%%%%%%%%%%
$$ i\{\Theta,\bar\Theta\} = \phi\bar\phi + c\bar{c}
- i c\bar{c}(\phi\bar\gamma - \gamma\bar\phi)
+ {3 \over 4}\gamma\bar\gamma c^2\bar{c}^2.\eqn\Laplacian$$
%%%%%%%%%%%%%%%%%%%%%%%%%%%%%%%%%%%%%%%%%%%%%%%%%%%%%%%%%%%%%%%%%%%%%%%%%%%%%
When $\gamma = 0$, this is simply the $OSp(1,\!1\,|\,2)$-invariant form.
We can prove the relation similar to \deltafcnI,
%%%%%%%%%%%%%%%%%%%%%%%%%%%%%%%%%%%%%%%%%%%%%%%%%%%%%%%%%%%%%%%%%%%%%%%%%%%%%
$$ \lim_{\beta\to\infty}\exp(-\beta i\{\Theta,\bar\Theta\}) = -\pi
\delta(\phi)\delta(\bar\phi)\delta(c)\delta(\bar{c}).\eqn\deltafcnII$$
%%%%%%%%%%%%%%%%%%%%%%%%%%%%%%%%%%%%%%%%%%%%%%%%%%%%%%%%%%%%%%%%%%%%%%%%%%%%%
To prove \deltafcnII, we integrate the left side against a test
function $\varphi(c,\bar{c})$, scale the ghosts and take the limit outside
the integral,
%%%%%%%%%%%%%%%%%%%%%%%%%%%%%%%%%%%%%%%%%%%%%%%%%%%%%%%%%%%%%%%%%%%%%%%%%%%%%
$$\eqalign{ &\lim_{\beta\to\infty}\int dc\,d\bar{c}\,\,\varphi(c,\bar{c})\,
\exp(-\beta i\{\Theta,\bar\Theta\}) \cr
= &\lim_{\beta\to\infty}\int{dc'\,d\bar{c'}\over\beta}\,\,
\varphi\Bigl({c'\over\sqrt\beta},{\bar{c'}\over\sqrt\beta}\Bigr)\,
e^{-c'\bar{c'}}\Bigl(1-\beta\phi\bar\phi\Bigr)\cr
&\phantom{\lim_{\beta\to\infty}\int{dc'\,d\bar{c'}\over\beta}}\,\,
\times\Bigl(1+i\, c'\bar{c'}(\phi\bar\gamma-\gamma\bar\phi)
- {3 \over 4\beta}(c'\bar{c'})^2\gamma\bar\gamma
- (c'\bar{c'})^2\phi\bar\gamma\gamma\bar\phi\Bigr)\cr
= &-\pi\,\phi\bar\phi\,\varphi(0,0).\cr}\eqn\deltaproof$$
%%%%%%%%%%%%%%%%%%%%%%%%%%%%%%%%%%%%%%%%%%%%%%%%%%%%%%%%%%%%%%%%%%%%%%%%%%%%%
We prove the general case of \deltafcnI\ in the next section.

\chapter{ The General Case }

\def\sumprime{{\sum_{n,m}\strut^\prime}}
\def\sumprimekl{{\sum_{k,\ell}\strut^\prime}} \def\xibar{{\overline{\xi}}}
\def\cbar{{\overline{c}}} \def\ibar{{\bar{\imath}}} \def\jbar{{\bar{\jmath}}}
\def\Ibar{{\bar I}} \def\Jbar{{\bar J}}
\def\cnbar#1{{\overline{c}\strut^{\,#1}}} \def\cn#1{{c\strut^{\,#1}}}
\def\xinbar#1{{\overline{\xi}\strut^{\,#1}}}  \def\xin#1{{\xi\strut^{\,#1}}}
\def\Xnm{X_{n,m}} \def\Xnmbar{\overline{X}_{n,m}} \def\Xinm{\Xi_{n,m}}
\def\Xinmbar{\overline{\Xi}_{n,m}}  
\def\Xklbar{\overline{X}_{k,\ell}}  
\def\Xiklbar{\overline{\Xi}_{k,\ell}}  \def\phibar{{\overline{\phi}}}  We now
consider a general constrained system with $N$ fermionic and $M$ bosonic
constraints satisfying the assumption of the method.  The general form of the
BRST charge $\Theta$ is
%%%%%%%%%%%%%%%%%%%%%%%%%%%%%%%%%%%%%%%%%%%%%%%%%%%%%%%%%%%%%%%%%%%%%%%%%%%%%
$$\eqalign{\Theta =& c\strut_I\phi\strut_I + \xi_i a_i + \hskip -18pt \sum_{n=0
\atop \phantom{n+n \geq 1} n+m \geq 1}^\infty \hskip -22pt \sum_{m=0 \atop
\phantom{n+m \geq 1}}^M  \cbar\strut_{\Ibar_1}\cdots\cbar\strut_{\Ibar_n}
c\strut_{I_1}\cdots c\strut_{I_n} \xibar_{\ibar_1}\cdots\xibar_{\ibar_m}
\xi_{i_1}\cdots\xi_{i_m}\cr  &\times\left(c\strut_{I_{n+1}}\Xi^{
\Ibar_1\cdots\Ibar_n I_1\cdots I_{n+1}\ibar_1\cdots\ibar_m i_1\cdots i_m} +
\xi_{i_{m+1}} X^{\Ibar_1\cdots\Ibar_n I_1\cdots I_n\ibar_1\cdots\ibar_m
i_1\cdots i_{m+1}} \right),\cr}\eqn\GenTheta$$
%%%%%%%%%%%%%%%%%%%%%%%%%%%%%%%%%%%%%%%%%%%%%%%%%%%%%%%%%%%%%%%%%%%%%%%%%%%%%
where the indices run $I=1,\ldots, N$ and $i=1,\ldots, M$. We write this
schematically as
%%%%%%%%%%%%%%%%%%%%%%%%%%%%%%%%%%%%%%%%%%%%%%%%%%%%%%%%%%%%%%%%%%%%%%%%%%%%%
$$\Theta = c\phi + \xi a +
\sum_{n,m}\strut^\prime\cnbar{n}\cn{n+1}\xinbar{m}\xin{m} \Xinm +
\sum_{n,m}\strut^\prime\cnbar{n}\cn{n}\xinbar{m}\xin{m+1}\Xnm,\eqn\Scheme$$
%%%%%%%%%%%%%%%%%%%%%%%%%%%%%%%%%%%%%%%%%%%%%%%%%%%%%%%%%%%%%%%%%%%%%%%%%%%%%
where $c^n$, for instance, denotes $c_{I_1}c_{I_2}\cdots c_{I_n}$ and the
primed sum $\sum^\prime_{n,m}$ is a sum  on all multi-indices of positive
length, $\{n+m\geq 1,\> n\geq 0, \> M\geq m \geq 0 \}$.
{}From \GenTheta\ it follows that the general BRST Laplacian is
%%%%%%%%%%%%%%%%%%%%%%%%%%%%%%%%%%%%%%%%%%%%%%%%%%%%%%%%%%%%%%%%%%%%%%%%%%%%%
$$\eqalign{i\{\Theta,\overline{\Theta}\} &= a_i\overline{a}_\ibar +
\phi_I\phibar_\Ibar + i\,\xi_i\{a_i, \overline{a}_\jbar\}\xibar_\jbar
+ i\,c_I\{\phi_I,\phibar_\Jbar\}\cbar_\Jbar
+ i\,\xi_i\{a_i,\phibar_\Jbar\}\cbar_\Jbar
+ i\,c_I\{\phi_I,\overline{a}_\jbar\}\xibar_\jbar \cr
&\quad + {\rm terms\>quadratic\> in\> ghosts,\>times\>constraints}\cr
&\quad + {\rm terms\>with\>more\>than\>
two\>ghosts},\cr}\eqn\Lapl$$
%%%%%%%%%%%%%%%%%%%%%%%%%%%%%%%%%%%%%%%%%%%%%%%%%%%%%%%%%%%%%%%%%%%%%%%%%%%%%
which we rewrite using DeWitt's supermatrix notation,\Ref\DeWitt{Bryce DeWitt,
{\it Supermanifolds\/}, (Cambridge University Press, Cambridge, 1992) p.20.}
%%%%%%%%%%%%%%%%%%%%%%%%%%%%%%%%%%%%%%%%%%%%%%%%%%%%%%%%%%%%%%%%%%%%%%%%%%%%%
$$i\{\Theta,\overline{\Theta}\} = a\overline{a} + \phi\phibar +
i\xi{\ss A}\xibar + ic{\ss B}\cbar +
i\xi{\ss C}\cbar + ic{\ss D}\xibar + \tilde\Delta. \eqn\LaplAbs$$
%%%%%%%%%%%%%%%%%%%%%%%%%%%%%%%%%%%%%%%%%%%%%%%%%%%%%%%%%%%%%%%%%%%%%%%%%%%%%
The last term, $\tilde\Delta$, contains all of the higher-order pieces.

To make the calculation of
$e^{-\beta i\{\Theta,\overline{\Theta}\}}$ in the general case similar to
the case of a single fermionic constraint considered above, it is useful to
rescale the bosonic and fermionic constraints and ghosts,
%%%%%%%%%%%%%%%%%%%%%%%%%%%%%%%%%%%%%%%%%%%%%%%%%%%%%%%%%%%%%%%%%%%%%%%%%%%%%
$$\eqalign{ a &= a'/\sqrt{\beta},\quad \bar{a} = \bar{a}'/\sqrt{\beta},
\quad c = c'/\sqrt{\beta},\quad \cbar = \cbar'/\sqrt{\beta}\cr
\phi &= \phi'/\sqrt{\beta},\quad \phibar = \phibar'/\sqrt{\beta},\quad
\xi = \xi'/\sqrt{\beta}, \quad\xibar= \xibar'/\sqrt{\beta}.\cr}
\eqn\primeghosts$$
%%%%%%%%%%%%%%%%%%%%%%%%%%%%%%%%%%%%%%%%%%%%%%%%%%%%%%%%%%%%%%%%%%%%%%%%%%%%%
We find
%%%%%%%%%%%%%%%%%%%%%%%%%%%%%%%%%%%%%%%%%%%%%%%%%%%%%%%%%%%%%%%%%%%%%%%%%%%%%
$$e^{-\beta i\{\Theta,\overline{\Theta}\}}=\exp\left( -\beta\phi\phibar - \beta
a\overline{a} - i\,\xi'{\ss A}\xibar' - i\,c'{\ss B}\cbar'  - i\,\xi'{\ss
C}\cbar' - i\,c'{\ss D}\xibar' - \beta\tilde\Delta \right), \eqn\rescale$$
%{\cal O}(\beta^{-{1\over2}})
%%%%%%%%%%%%%%%%%%%%%%%%%%%%%%%%%%%%%%%%%%%%%%%%%%%%%%%%%%%%%%%%%%%%%%%%%%%%%
which we rewrite in the suggestive form
%%%%%%%%%%%%%%%%%%%%%%%%%%%%%%%%%%%%%%%%%%%%%%%%%%%%%%%%%%%%%%%%%%%%%%%%%%%%%
$$e^{-\beta i\{\Theta,\overline{\Theta}\}}=\exp\left( -\beta\phi\phibar - \beta
a\overline{a} - i\,\xi'({\ss A -CB^{-1}D})\xibar'  - i\,(c'+\xi'{\ss
CB^{-1}}){\ss B}(\cbar' + {\ss B^{-1}D}\xibar')  - \beta\tilde\Delta
\right).\eqn\shift$$ %{\cal O}(\beta^{-{1\over2}})
%%%%%%%%%%%%%%%%%%%%%%%%%%%%%%%%%%%%%%%%%%%%%%%%%%%%%%%%%%%%%%%%%%%%%%%%%%%%%
The last piece, $\beta\tilde\Delta$, is the sum of the terms,
$\beta\tilde\Delta = \beta\tilde\Delta^{(1)} + \beta\tilde\Delta^{(2)}$,
\begingroup \def\PB#1{{\{#1\} }} \def\cc#1#2{{\,\overline{c}^{\prime\, #1}\,
c^{\prime\, #2}\,}}
\def\xixi#1#2{{\overline{\xi}'\strut^{\,#1}\,\xi'\strut^{#2}\,}}
$$\eqalign{\beta\tilde\Delta^{(1)} = \sumprime
&\Bigl(a'\cc{n+1}{n}\xixi{m-1}{m}\Xinmbar\,\,\beta^{-n-m+{1\over2}}\cr &+
i\,\cc{n+1}{n}\xixi{m}{m+1}\PB{a,\Xinmbar}\,\,\beta^{-n-m}\cr &+
a'\cc{n}{n}\xixi{m}{m}\Xnmbar\,\,\beta^{-n-m+{1\over2}}\cr &+
i\,\cc{n}{n}\xixi{m+1}{m+1}\PB{a,\Xnmbar}\,\,\beta^{-n-m}\cr &+ \phi' \cc{n}{n}
\xixi{m}{m}\Xinmbar\,\,\beta^{-n-m+{1\over2}}\cr &+ i\,\cc{n+1}{n+1}
\xixi{m}{m}\PB{\phi,\Xinmbar}\,\,\beta^{-n-m}\cr &+ \phi'
\cc{n-1}{n}\xixi{m+1}{m}\Xnmbar\,\,\beta^{-n-m+{1\over2}}\cr &+
i\,\cc{n}{n+1}\xixi{m+1}{m} \PB{\phi,\Xnmbar}\,\,\beta^{-n-m}\Bigr) \cr
&+c.c.\cr}\eqn\Deltaone$$ $$\eqalign{\beta\tilde\Delta^{(2)} =
\sumprime\sumprimekl &\Bigl(
\cc{n+k}{n+k}\xixi{\ell+m}{\ell+m}\Xinm\Xiklbar\,\,\beta^{1-k-n-\ell-m}\cr
&+\cc{n+k+1}{n+k+1}\xixi{\ell+m-1}{\ell+m-1}\Xinm\Xiklbar\,\,
\beta^{1-k-n-\ell-m}\cr
&+i\,\cc{n+k+1}{n+k+1}\xixi{\ell+m}{\ell+m}\PB{\Xinm,\Xiklbar}\,\,
\beta^{-k-n-\ell-m}\cr &+\cc{n+k-1}{n+k}\xixi{\ell+m+1}{\ell+m}\Xinm\Xklbar\,\,
\beta^{1-k-n-\ell-m} +c.c.\cr
&+\cc{n+k}{n+k+1}\xixi{\ell+m}{\ell+m-1}\Xinm\Xklbar\,\, \beta^{1-k-n-\ell-m}
+c.c.\cr &+i\,\cc{n+k}{n+k+1}\xixi{\ell+m+1}{\ell+m}\PB{\Xinm,\Xklbar}\,\,
\beta^{-k-n-\ell-m} +c.c.\cr
&+\cc{n+k-1}{n+k-1}\xixi{\ell+m+1}{\ell+m+1}\Xnm\Xklbar\,\,
\beta^{1-n-k-\ell-m}\cr &+\cc{n+k}{n+k}\xixi{\ell +m}{\ell
+m}\Xnm\Xklbar\,\,\beta^{1-n-k-\ell-m}\cr
&+i\,\cc{n+k}{n+k}\xixi{\ell+m+1}{\ell+m+1}\PB{\Xnm,\Xklbar}\,\,
\beta^{-n-k-\ell-m} \Bigr) \cr}\eqn\Deltatwo$$
%%%%%%%%%%%%%%%%%%%%%%%%%%%%%%%%%%%%%%%%%%%%%%%%%%%%%%%%%%%%%%%%%%%%%%%%%%%%%
\endgroup

The terms in \Deltaone\ and \Deltatwo\ above are all at least
${\cal O} (\beta^{-{1\over2}})$, which means that they can be ignored in the
limit $\beta\to\infty$. To obtain the delta function relation analogous to
the Senjanovic measure \deltafcnI, we integrate
against a test function $\varphi$ of the bosonic variables
$c,\cbar,a,\overline{a}$.  The integral we wish to evaluate is
%%%%%%%%%%%%%%%%%%%%%%%%%%%%%%%%%%%%%%%%%%%%%%%%%%%%%%%%%%%%%%%%%%%%%%%%%%%%%
$$\eqalign{ I_\beta=\int&
d^N\!c\,\,d^N\!\cbar\,\,d^M\!a\,\,d^M\!\overline{a}\,\, e^{-\beta
i\{\Theta,\overline{\Theta}\}}\,\varphi(c,\cbar,a,\overline{a})\cr
=\int&{d^N\!c'\,\,d^N\!\cbar'\,\,d^M\!a'\,\,d^M\!\overline{a}'\over
\beta^{N+M}} \,\,  \varphi\left({c'\over\sqrt\beta},{\cbar'\over\sqrt\beta},
{a'\over\sqrt\beta},{\overline{a}'\over\sqrt\beta} \right)\cr &\times
e^{-\left(\beta\phi\phibar + a'\overline{a}' + i\xi'({\ss A -CB^{-1}D})\xibar'
+ i(c'+\xi'{\ss CB^{-1}}) {\ss B}(\cbar' + {\ss B^{-1}D}\xibar') + {\cal
O}(\beta^{-{1\over2}})\right)},\cr}\eqn\testeval$$
%%%%%%%%%%%%%%%%%%%%%%%%%%%%%%%%%%%%%%%%%%%%%%%%%%%%%%%%%%%%%%%%%%%%%%%%%%%%%
which, upon a shift of the ghost variables, $c'\to c' -\xi'{\ss CB}^{-1}$,
becomes
%%%%%%%%%%%%%%%%%%%%%%%%%%%%%%%%%%%%%%%%%%%%%%%%%%%%%%%%%%%%%%%%%%%%%%%%%%%%%
$$\eqalign{I_\beta=(-1)^{N+M}\int&{d^N\!c'\,\,d^N\!\cbar'\,\,d^M\!a'\,\,d^M\!
\overline{a}'\over\beta^{N+M}}\,\,\varphi\left({c'\over\sqrt\beta},
{\cbar'\over\sqrt\beta},{a'\over\sqrt\beta},{\overline{a}'\over\sqrt\beta}
\right)\beta^N\delta^N(\phi)\delta^N(\phibar)\cr &\times
\beta^M\delta^M(\xi)\delta^M(\xibar)\det\lbrack i({\ss A - CB^{-1}D}) \rbrack
e^{-a'\overline{a}'}\,e^{-ic'{\ss B}\cbar'}\,+{\cal
O}(\beta^{-{1\over2}}).\cr}\eqn\integral$$
%%%%%%%%%%%%%%%%%%%%%%%%%%%%%%%%%%%%%%%%%%%%%%%%%%%%%%%%%%%%%%%%%%%%%%%%%%%%%
Because $\varphi$ is a test function, we obtain in the limit
%%%%%%%%%%%%%%%%%%%%%%%%%%%%%%%%%%%%%%%%%%%%%%%%%%%%%%%%%%%%%%%%%%%%%%%%%%%%%
$$\lim_{\beta\to \infty}I_\beta = (-\pi)^{N+M} \delta^N(\phi)\delta^N(\phibar)
\delta^M(\xi)\delta^M(\xibar)\det\lbrack i({\ss A - CB^{-1}D})\rbrack
(\det(i{\ss B}))^{-1}\varphi({\bf 0},{\bf 0},{\bf 0},{\bf 0}),\eqn\limitrel$$
%%%%%%%%%%%%%%%%%%%%%%%%%%%%%%%%%%%%%%%%%%%%%%%%%%%%%%%%%%%%%%%%%%%%%%%%%%%%%
which proves the general case of \deltafcnI,
%%%%%%%%%%%%%%%%%%%%%%%%%%%%%%%%%%%%%%%%%%%%%%%%%%%%%%%%%%%%%%%%%%%%%%%%%%%%%
$$\eqalign{\lim_{\beta\to \infty}e^{-\beta i\{\Theta,\overline{\Theta}\}}
&=(-\pi)^{N+M} \delta^N(\phi_I)\delta^N(\phibar_\Jbar)\delta^M(\xi_i)
\delta^M(\xibar_\jbar)\, {\rm sdet}\pmatrix{i\{a_i, \overline{a}_\jbar\}&
i\{a_i,\phibar_\Jbar\}\cr i\{\phi_I,\overline{a}_\jbar\}&
i\{\phi_I,\phibar_\Jbar\}\cr}\cr  &\phantom{=}\times \delta^N(\Re
c_I)\delta^N(\Im \cbar_\Jbar)\delta^M(\Re a_i)\delta^M(\Im
\overline{a}_\jbar).\cr} \eqn\mainresult$$
%%%%%%%%%%%%%%%%%%%%%%%%%%%%%%%%%%%%%%%%%%%%%%%%%%%%%%%%%%%%%%%%%%%%%%%%%%%%%

In summary, our main result is that the harmonic BRST-BFV method  introduced in
ref.\ [\HBRST] generalizes to the case of mixed bosonic and fermionic
constraints.  There is no problem in extending the operator formalism, but it
is not trivial to show that the extension also works for the path integral. It
is possible that the relation \mainresult\ is only valid for test functions and
not for rapidly decreasing functions as well.   This is because  the bosonic
part of the exponent ${-\beta i\{\Theta,\overline{\Theta}\}}$ might not be
negative definite, although the quadratic piece by itself is negative definite.
For test functions, this is irrelevant because in the limit the quadratic piece
dominates all others on any finite interval.  In the case of purely bosonic
constraints, this is not an issue because the higher order pieces are fermionic
and have no convergence problem.   To be rigorous about a particular path
integral, one must check in that specific case that the result also holds for
functions of rapid decrease.

Our modification of the path integral and our limit argument are very similar
to the method of equivariant localization recently introduced by Dykstra,
Lykken and Raiten.\Ref\DyLyRa{H.M. Dykstra, J.D. Lykken and E.J. Raiten, {\it
``Exact Path Integrals by Equivariant Localization,"\/} FERMI-PUB-92/383-T,
UMHEP-384, hep-th preprint 9212126.} Invariance under the change of variables
generated by the holomorphic BRST charge $\Theta$ in our formalism corresponds
closely to the invariance under the equivariant exterior derivative $d_\chi$ in
the formalism of ref.\ [\DyLyRa].

The extension of the formalism to the reducible case is an open problem, but
one whose solution is quite likely to follow the standard reducible BRST
quantization for real constraints.

\ack              % this generates the acknowledgement heading

It is a pleasure to thank Bernice Durand for her careful reading of this paper
and her many useful suggestions.  This work was supported in part by DOE grant
No.\ DE-AC02-76-ER00881.

\refout           % this outputs the reference page
\bye